\documentclass[aps,pre,twocolumn,showpacs,preprintnumbers,amsmath,amssym,superscriptaddress]{revtex4}
\usepackage{graphicx}
\usepackage{bm,color,textcomp}
\usepackage{amsmath,amsfonts,amssymb}

\usepackage{subfig}
\usepackage{graphicx}

\usepackage[normalem]{ulem}

\begin{document}

\title{Excess and loss of entropy production for different levels of coarse-graining}
\author{Pierpaolo Bilotto}
\affiliation{Dipartimento di Fisica, Universit\'a di Roma Sapienza, P.le Aldo Moro 5, 00185, Rome, Italy}
\author{Lorenzo Caprini}
\email{lorenzo.caprini@gssi.it}
\affiliation{Scuola di Scienze e Tecnologie, Universit\'a di Camerino - via Madonna delle Carceri, 62032, Camerino, Italy}
\author{Angelo Vulpiani}
\affiliation{Dipartimento di Fisica, Universit\'a di Roma Sapienza, P.le Aldo Moro 5, 00185, Rome, Italy}

\date{\today}

\begin{abstract}
We investigate the effect of coarse-graining on the energetics properties of a system, focusing on entropy production.
As a case of study, we consider a one-dimensional colloidal particle in contact with a thermal bath, moving in a sinusoidal potential and driven out of equilibrium by a small constant force. Different levels of coarse-graining are evaluated: at first, we compare the results in the underdamped dynamics with those in the overdamped one (first coarse-graining).
For large values of the viscosity, the two dynamics have the same energetic properties, while, for smaller viscosity values, the overdamped approximation produces an excess of entropy production with respect to that of the underdamped dynamics.
Moreover, for further smaller values of the drag coefficient, the excess of entropy production turns into a loss.
These regimes are explained by evaluating the jump statistics, observing that the inertia is able to induce multiple jumps and affect the average jump rate.
The periodic shape of the potential allows us to approximate the continuous dynamics via a Markov chain, after the introduction of a suitable time and space discretization (second level of coarse-graining).
This discretization procedure is implemented starting both from the underdamped and the overdamped evolution and is analyzed for different values of the viscosity.
\end{abstract}

\maketitle

\section{Introduction}\label{sec:intro}


Physical and biological systems often involve a plethora of different processes, each described by a suitable variable and characterized by a typical time-scale.
In a broad range of interesting cases, the typical times of the different variables are much different and the system is said to have a multiscale character, i.e. some variables are faster than the global timescale of the dynamics, which is much slower.
For instance, this is the case of protein folding where the timescale of vibration of covalent bonds is $O(10^{-15}) \text{s}$ while the folding time for a protein may be of order of seconds~\cite{baldwin2000structure, finkelstein2016protein}. 
Therefore, one faces with the problem of understanding the most relevant features of the system that often corresponds to treating the ``slow part of the dynamics'' in terms of effective equations, performing the so-called coarse-graining~\cite{kadanoff2000statistical, castiglione2008chaos, saunders2013coarse}.
Such a necessity is practical (even modern computers  sometimes are not able to simulate all the relevant scales involved in certain problems) as well as conceptual: effective equations are able to catch some general features and to evidence basic ingredients which can be hidden in the detailed description.
 
In the practical research activity the use of different levels of description, or in other words different coarse-graining procedures, is unavoidable.
The case of mesoscopic objects, such as a colloids immersed in solution, illustrate this point: in principle, one can adopt a very detailed microscopic description in terms of Hamiltonian dynamics, so that the evolution appears reversible in time but involves a large number of degrees of freedom.
To overcome numerical and theoretical difficulty, the most common approach neglects some "details" of the system (namely, the dynamics of the solvent molecules) in favor of a mesoscopic description in terms of random variables: this corresponds to a temporal and spatial coarse-graining.
Since one can introduce different possible mesoscopic descriptions, usually different levels of coarse-graining are allowed.
Therefore, evaluating the consequences of the coarse-graining on the physical properties of the system is a general issue of indisputable importance. For this reason, their effects on the energetic properties, such as entropy production of the system, have been the focus of intense research~\cite{seiferth2020coarse, sohn2015example, teza2020exact, lucarini2014entropy, chun2015hidden} performing both spatial and time discretizations of continuous dynamics and constructing the transition rates of suitable Markov chains~\cite{puglisi2010entropy, knoch2019non, sohn2016critical}.
For instance, Busiello et al. evaluate the difference between the entropy production of a discrete multi-body process and an approximate description based on a continuous dynamics~\cite{busiello2019entropy, busiello2019entropy1}, while the authors of Ref.~\cite{crisanti2012nonequilibrium} study the entropy production of a continuous process accounting (and not) for the faster degree of freedom and focusing on the role of cross correlations.
While well-separated time-scales allow to calculate the entropy production neglecting the fast degree of freedom, this is not anymore true when temperature spatial gradients are included in the dynamics \cite{celani2012anomalous}.

In this paper, we consider as a reference system a particle in contact with a thermal bath moving in a tilted potential, 
which has been employed to describe the motion of molecular motors, enzyme turnover reactions~\cite{kolomeisky2007molecular, astumian2010thermodynamics, hwang2017quantifying}, and the diffusion of a colloidal particle in a rotating array of laser traps \cite{evstigneev2008diffusion}. In addition, a two-layer colloidal system has been implemented to study experimentally the diffusion of a colloid over a periodic potential \cite{ma2013colloidal, ma2015colloidal}.
The energetic properties of this system, such as heat rate and entropy production, have been analyzed in some previous studies, for overdamped~\cite{speck2007distribution, hyeon2017physical} or underdamped dynamics~\cite{kawaguchi2013fluctuation, fischer2018large, taye2020entropy} often focusing on critical regimes of the driving force~\cite{gopal2021energetics}.
Here, we evaluate the effect of coarse-graining on the energetic properties of the system, e.g. entropy production, for increasing levels of coarse-graining by exploring both approximate continuous and discrete dynamics. 

The paper is structured as follows:
in Sec.~\ref{sec:model}, we introduce the underdamped dynamics 
as a reference model and the first level of coarse-graining, namely the overamped description.
The expressions for the entropy production are reported in both cases while their derivations are reviewed in the Appendices.
The numerical study of the entropy production is shown in Sec.~\ref{sec:results} for both the dynamics, together with the study of the jump statistics.
In Sec.~\ref{sec:discrete}, the second level of coarse-graining based on the space and time-discretization is introduced based on the master equation and suitable Markov chain dynamics. 
In Sec.~\ref{sec:numerical_study_discrete}, the entropy production resulting from the discrete coarse-graining is numerically evaluated and compared with the result of the continuous evolution.
Finally, we summarize and discuss the results in the conclusions presenting some future perspectives.

\section{Model}
\label{sec:model}

\begin{figure}[!t]
\centering
\includegraphics[width=0.95\linewidth,keepaspectratio]{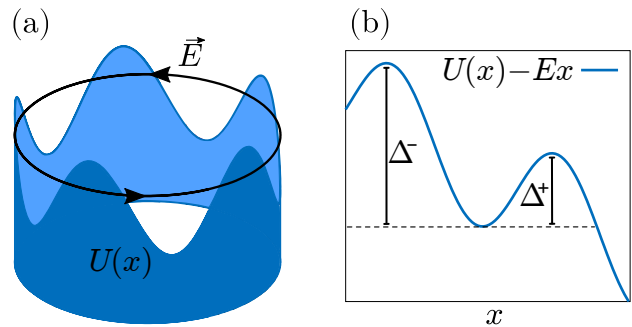}
\caption{Periodic potential. In panel (a), a three-dimensional representation of the periodic potential $U(x)$ is shown while, in 
panel (b), the whole potential profile $U(x)-Ex$ is reported, where $U(x)=A \cos(2 \pi x/L)$. 
Vertical lines in panel (b) draw the heigthes of the potential barriers for forward and backward jumps, $\Delta^+$ and $\Delta^-$, respectively. 
The parameters of the potential are $A=4$ and $E=1$.
}
\label{fig:potential}
\end{figure}

We consider the dynamics of a one-dimensional particle in contact with a thermal bath at temperature, $T$, moving in a periodic potential of the form
$$
U(x)=A \cos(2\pi x /L) \,,
$$
where $A$ determines the amplitude of the oscillation and $L$ the distance between two potential minima.
The particle is driven out-of-equilibrium by a constant force, $E>0$, pointing along with $\hat{x}$, that produces a net current in that direction.
The asymmetry introduced by the driving force changes the effective height of the potential barriers (tilted potentials), from $2A$ to $\Delta^{\pm}$, breaking the left-right spatial symmetry and introducing a preferential direction of motion, as illustrated In Fig.\ref{fig:potential} (b):
the jumps in the same direction of the force (forward jumps) are facilitated, being $\Delta^+ < 2A$, the jumps in the opposite direction (backward jumps) are hindered because $\Delta^{-}>2A$.
The value of $\Delta^{\pm}$ can be calculated and reads:
$$
\frac{\Delta^\pm}{E\,L}=\frac{\sqrt{a^2-1}+\sin^{-1}(1/a)}{\pi}\pm\frac{1}{2} \,,
$$
where $a=2\pi A/(E L)$ quantifies the strength of the driving force with respect to the amplitude of the periodic potential.
If $E$ approaches the critical value, i.e. $E\geq E_c=2\pi A/L$ the jump mechanism is entirely due to the driving force and the trajectory is almost deterministic.
In what follows, we restrict our analysis to the regime $E\ll E_c$, so that $E$ is a small perturbation with respect to the force due to $U(x)$ and, thus, in principle, it is not able to suppress the spatial oscillating structure of the potential.
In other words, the jumps from a potential minimum to the others should occur only because of thermal fluctuations.

\subsection{The underdamped dynamics}

The reference description that is employed to describe the particle motion is the underdamped Langevin equation.
This is taken as the lowest level of coarse-graining and, thus, as the reference case for the {\it exact} evolution.
At this level of description, the dynamics for the position, $x$, and the velocity, $v$, of a particle with mass, $m$, is described by the following stochastic equations
\begin{subequations}
\label{eq:underdamped_dynamics}
\begin{align}
\dot{x}&=v \,, \\
m \dot{v}&=-\gamma v  + \sqrt{2 T \gamma} \xi+ U'(x)  + E \,,
\end{align}
\end{subequations}
where $\gamma$ and $T$ are the solvent viscosity and temperature of the bath, respectively, that satify the Einstein's relation and
$\xi$ is a white noise with zero average and $\langle\xi(t) \xi(0) \rangle = \delta(t)$. The physical origin of the noise is injection of energy due to the fast collisions of the solvent molecules with the tagged particle. 
As usual, this energy injection is balanced by the Stokes force, proportional to $v$ via $\gamma$, that determines the relaxation time of the velocity, namely the inertial time $\tau_I =m /\gamma$. 

Using path-integral or alternative methods based on the distinction between reversible and irreversible currents, it is possible to derive an explicit expression for the rate of entropy production \cite{seifert2012stochastic, spinney2012entropy, peliti2021stochastic}, $\dot{\sigma}$, associated to the dynamics~\eqref{eq:underdamped_dynamics}: 
\begin{equation}
\label{eq:sigma_underdamped}
\dot{\sigma}=\frac{E}{T}\langle v\rangle \,,
\end{equation}
where $\langle \cdot \rangle$ is the steady-state average to be performed with the steady-state measure of the dynamics~\eqref{eq:underdamped_dynamics}.
In the expression~\eqref{eq:sigma_underdamped}, we have neglected boundary terms because they are irrelevant in the steady-state. 

\subsection{First level of coarse-graining: overdamped dynamics}

The first level of approximation for the dynamics~\eqref{eq:underdamped_dynamics} consists in neglecting the inertial forces.  
This procedure is well-justified in the limit of small inertial time, $m/\gamma$, where the velocity relaxation occurs fastly and, therefore, this variable can be adiabatically eliminated.
Within this approximation, we introduce the first level of coarse-graining, where the particle is uniquely described by a stochastic evolution for its position $x$:
\begin{equation}
\label{eq:overdamped_dynamics}
\gamma \dot{x}  = \sqrt{2 T \gamma} \xi+ U'(x)  + E \,.
\end{equation}
The dynamics~\eqref{eq:overdamped_dynamics} is much simpler than Eq.~\eqref{eq:underdamped_dynamics} and allows one to derive some analytical results, which are unknown in the underdamped case.
For instance, the jump time, $\tau_o^{\pm}$, from a well to the foward or backward one, respectively, can be analytically predicted in the limit of small temperature, such that $T/\Delta^{\pm}\ll1$, where the well-known Kramers formula holds:
\begin{equation}
\label{eq:Kramerstime}
\tau_o^{\pm} \approx \frac{L \gamma}{E\sqrt{a^2-1}} \exp{\left(\frac{\Delta^{\pm}}{T}\right)} \,,
\end{equation}
where $a>1$, as occurs in the regime of small driving forces studied in this paper.

Also in the overdamped case, the expression for the rate of entropy production, $\dot{\sigma}_o$, of a system following the dynamics~\eqref{eq:overdamped_dynamics} can be obtained \cite{speck2007distribution}: 
\begin{equation}
\label{eq:sigma_overdamped}
\dot{\sigma}_o=\frac{E}{T}\langle \dot{x}\rangle_o \,,
\end{equation}
where, now, $\langle \cdot\rangle_o$ is the steady-state average performed using the steady-state probability distribution associated with the dynamics~\eqref{eq:overdamped_dynamics} and $\dot{x}$ is the coarse-grained velocity of the overdamped system. 
Again, this formula express the entropy production rate apart from boundary terms, which are negligible in the steady-state.

\section{Entropy production: comparison between underdamped and overdamped dynamics}
\label{sec:results}

\begin{figure*}[!t]
\centering
\includegraphics[width=0.95\linewidth,keepaspectratio]{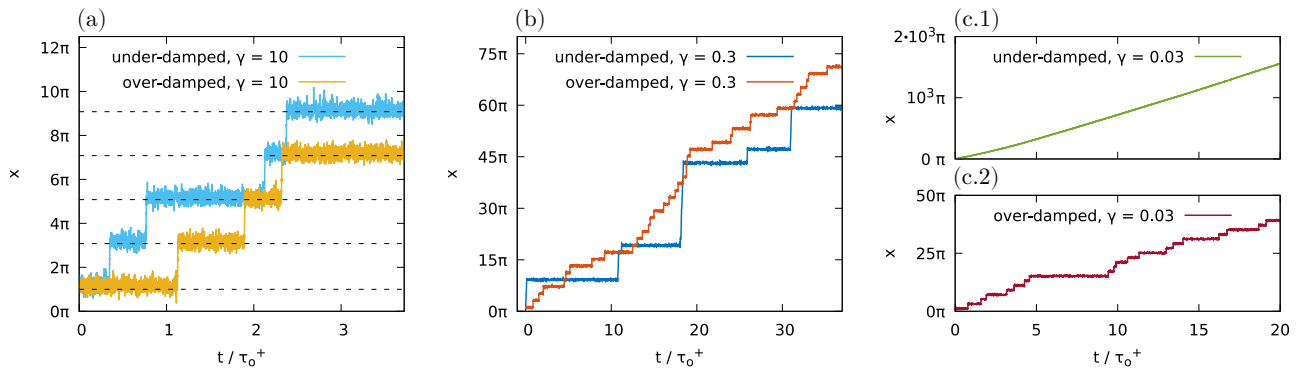}
\caption{Single-particle Trajectories.  Panels (a) and (b)  and (c) show the time-trajectory of the position $x(t)$ for three different values of $\gamma=10, 3\times 10^{-1}, 3 \times 10^{-2}$, respectively. In each panels, the underdamped (simulating Eq.~\eqref{eq:underdamped_dynamics}) and the overdamped cases (simulating Eq.~\eqref{eq:overdamped_dynamics}) are shown (the two cases are separately plotted in two different plots, namely (c.1) and (c.2) for $\gamma=3\times10^{-2}$ for presentation reasons.
The other parameters of the numerical study are: $L=2\pi$, $E=1$, $A=4$, $T=1$ and $m=1$.
}
\label{fig:traj}
\end{figure*}

\begin{figure}[!t]
\centering
\includegraphics[width=0.95\linewidth,keepaspectratio]{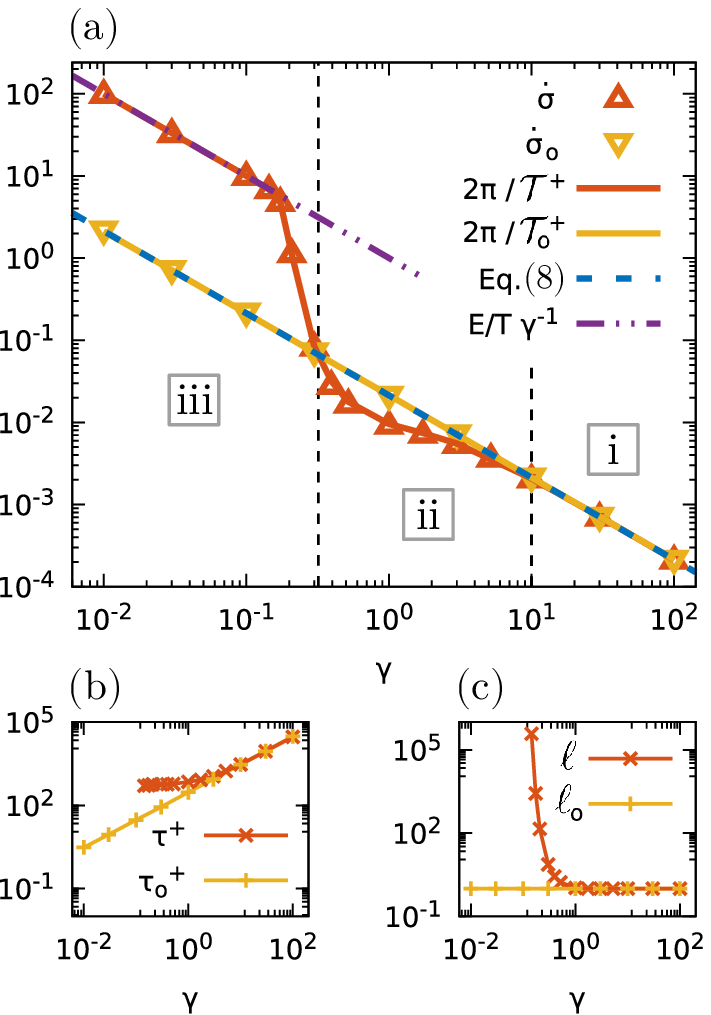}
\caption{Coarse-graining i): underdamped $\to$ overdamped. Panel (a): entropy productions, $\dot{\sigma}$ (red upper triangles) and $\dot{\sigma}_o$ (yellow lower triangles), as a function of the viscosity $\gamma$. 
$\dot{\sigma}$ is calculated numerically by using Eq.~\eqref{eq:sigma_underdamped} while $\dot{\sigma}_o$ by Eq.~\eqref{eq:sigma_overdamped}, corresponding to the underdamped and the overdamped cases, respectively. 
Solid colored lines are obtained by plotting $2\pi/\mathcal{T}$ and $2\pi/\mathcal{T}_o$ for the underdamped and the overdamped cases. 
Finally, the prediction~\eqref{eq:prediction_sigmao} is reported as a blue dashed line.
Panel (b) plots the average jump time from , $\tau_+$ (underdamped) and $\tau^+_o$ (overdamped) as a function of $\gamma$, while panel (c) plots the average length in units of $L$ run in each jump process, namely $\ell$ (underdamped) and $\ell_o$ (overdamped).
The other parameters of the simulations are: $L=2\pi$, $E=1$, $A=4$, $T=1$ and $m=1$.
}
\label{fig:continous_EP}
\end{figure}

The dynamics~\eqref{eq:underdamped_dynamics} and~\eqref{eq:overdamped_dynamics} have been numerically studied using the Heun's integration scheme.
In the numerical simulations, the shape of the potential is fixed by setting $a=A=4$ and $E=1$ and a regime of small temperature is considered so that one could expect jumps from a minimum to the others to be rare.
To unveil the effect of the inertial forces, we consider different inertial times by varying $\gamma$ and keeping the mass fixed ($m=1$).
Before delving into the study of entropy production, it is instructive to evaluate the single-particle trajectories to understand the mechanism behind the jump events.
Fig.~\ref{fig:traj} shows three typical trajectories, $x(t)$, both for the underdamped and overdamped dynamics, for three different values of $\gamma= 3\times10^{-2}, 3\times10^{-1}, 10$.
For $\gamma=10$, the two trajectories show almost indistinguishable statistical features displaying the jump mechanism illustrated by the single-particle trajectories in panel (a): the particle moves around a potential minimum and rarely (namely, after a typical time $\tau^+\approx\tau^+_o$) performs a forward jump (because the probability of jumping backward is negligible).
For $\gamma=3\times10^{-1}, 3\times10^{-2}$, the trajectory obtained integrating the dynamics~\eqref{eq:overdamped_dynamics} (overdamped case) shows a behavior qualitatively similar to that observed for $\gamma=10$, while a fairly different scenario occurs evolving the dynamics~\eqref{eq:underdamped_dynamics} (underdamped case): after spending a considerable large time in a minimum of the potential, the particle usually performs multiple jumps as clearly shown in panel (b);
instead, for $\gamma= 3\times10^{-2}$, the particle moves almost deterministically without fluctuating around the potential minima and, in practice, behaves as if $E>E_c$.
As a consequence, we conclude that the main effect of the inertial forces is to affect the jump mechanism, at first producing multiple jumps and, for further large inertial times, even to decrease the effective critical value of the driving force. 

\subsection{Entropy production}

In order to understand the effect of the coarse-graining on the energetics of the system, In Fig.~\ref{fig:continous_EP}, we compare $\dot{\sigma}$ and $\dot{\sigma}_o$, given by Eq.~\eqref{eq:sigma_underdamped} and Eq.\eqref{eq:sigma_overdamped}, respectively, for different values of $\gamma$ (at fixed $m=1$), exploring regimes characterized by large and small inertial times.
Even if $\dot{\sigma}_o$ formally coincides with $\dot{\sigma}$
the two expressions are calculated using two different steady-state averages and, thus, one cannot guarantee that $\dot{\sigma}_o$ remains a good approximation for $\dot{\sigma}$.
In both cases, the two entropy productions decrease monotonically when $\gamma$ is increased.
Specifically, $\dot{\sigma}_o$ decreases as $1/\gamma$ for the whole range of $\gamma$ values numerically analyzed.
Instead, the entropy production of the underdamped dynamics $\dot{\sigma}$ shows a richer behavior: i) for $\gamma>10$, $\dot{\sigma}$ decreases as $1/\gamma$ and coicides with $\dot{\sigma}_o$.
This result was expected because, in this regime of parameters, the adiabatic elimination of the velocity can be achieved and, thus, the underdamped steady-state distribution function of the position is well-reproduced by the overdamped one.
ii) For intermediate values of $\gamma$, namely for $3\times10^{-1}<\gamma<10$, $\dot{\sigma}_o$ overestimates $\dot{\sigma}$, showing a maximal discrepancy between the two descriptions for $\gamma \approx 1$.
iii) When $\gamma < 3\times 10^{-1}$, we get the opposite result since $\dot{\sigma}_o$ underestimates $\dot{\sigma}$. In particular, $\dot{\sigma}$ sharply increases until reaching another asymptotic regime that scales as $1/\gamma$ but its value is fairly different from that of $\dot{\sigma}_o$.
Regime iii) occurs in correspondence of the $\gamma$-values at which the particle starts behaving as if the driving force overcame its critical value (see Fig.~\ref{fig:traj}~(c)), i.e. when the jump process occurs deterministically.
In particular, in this regime, $\langle v \rangle$ roughly approaches the limiting value $\sim E/\gamma$ explaining the $1/\gamma$-scaling and the large value of the entropy production observed in Fig.~\ref{fig:continous_EP}.

The dependence of the entropy production on $\gamma$ can be understood by estimating $\langle v \rangle$ (or $\langle \dot{x} \rangle$) by the average time, $\mathcal{T}$ (or $\mathcal{T}_o$), needed to jump from a minimum of $U(x)$ to the forward one.
We remark that $\mathcal{T}$ and $\mathcal{T}_o$ correspond with $\tau^+$ and $\tau^+_o$, respectively (namely, the average jump times in underdamped and overdamped regimes, respectively) only in the regime when multiple jumps are negligible.
In this way, $\dot{\sigma}$ and $\dot{\sigma}_o$ can be estimated as:
\begin{flalign}
\label{eq:times_under}
&\dot{\sigma}\propto\langle v \rangle \approx \frac{L}{\mathcal{T}} \,\\
\label{eq:times_over}
&\dot{\sigma}_o\propto\langle \dot{x} \rangle \approx  \frac{L}{\mathcal{T}_o}\,.
\end{flalign}
Predictions~\eqref{eq:times_under} and~\eqref{eq:times_over} are numerically checked in Fig.~\ref{fig:continous_EP}~(a) (solid colored lines) by calculating $\mathcal{T}$ and $\mathcal{T}_o$ by simulations for each value of $\gamma$.
In the overdamped case, the linear dependence with $\gamma$ predicted by the Kramers formula~\eqref{eq:Kramerstime} is confirmed
and, in particular, the value of $\dot{\sigma}_o$ is also in fair agreement with its theoretical prediction (see the comparison between the solid and the dashed lines, again in Fig.~\ref{fig:continous_EP}~(a)):
\begin{equation}
\dot{\sigma}_o= \frac{2E}{\gamma L} \sinh\left(\frac{EL}{2T}\right) \left|I_{i\frac{EL}{2\pi T}}\left(\frac{A}{T}\right)\right|^{-2} \,,
\label{eq:prediction_sigmao}
\end{equation}
where $I$ is the modified Bessel function of the first kind.
The derivation of the prediction \eqref{eq:prediction_sigmao} can be found in Ref.~\cite{gopal2021energetics}.

In the underdamped case, the regimes i) and ii) and iii) and, in particular, the non-linear $\gamma$-dependence are observed also in the behavior of $1/\mathcal{T}$, showing that the entropy production is uniquely determined by the numbers of minima explored by the particle in a given interval of time.
To understand regime ii), we explore in more detail the jump mechanism: Fig.~\ref{fig:continous_EP}~(b) plots the average jump times, $\tau^+$ (underdamped) and $\tau^+_o$ (overdamped), as a function of $\gamma$, while Fig.~\ref{fig:continous_EP}~(c) measures the occurrence of multiple jumps in the system by plotting, as a function of $\gamma$, the average length, $\ell$ (in units of $L$) run in each jump process.
While for large values of $\gamma$ (i.e. regime i)), we have $\tau^+ \approx \tau^+_o \propto \gamma$ and $\ell \approx 1$ (i.e. absence of multiple jumps), the jump mechanism dramatically changes for intermediate values of $\gamma$ (i.e. regime ii)).
The first difference between $\dot{\sigma}$ and $\dot{\sigma}_o$ is mainly due to the increase of the average jump time $\tau^+$ with respect to $\tau^+_o$ (when multiple jumps are still negligible): when $\gamma$ decreases the correlation time of the velocity increases and, thus, a larger time is needed by the particle to explore the tail of the velocity distribution, i.e. the large values necessary to jump the barrier height.
For further smaller values of $\gamma$, this mechanism is balanced by the occurrence of multiple jumps (in agreement with the observation of the single-particle trajectory in Fig.~\ref{fig:traj}~(b)).
In particular, $\dot{\sigma}$ starts inceasing with respect to $\dot{\sigma}_o$ because when the $\tau_o^+$ saturates the $\ell$ sharply increases.

In conclusion, inertial effects are able to hinder or facilitate the forward jumps with respect to the overdamped case by affecting the average jump time and causing frequent multiple jumps. Consequently, the underdamped $\to$ overdamped coarse-graining procedure produces a loss or an excess of entropy production depending on the value of the viscous drag coefficient considered ($\sim$ inverse of the inertial time).

\section{Discrete coarse-graining}\label{sec:discrete}

\begin{figure*}
\centering
\includegraphics[width=0.9\linewidth,keepaspectratio]{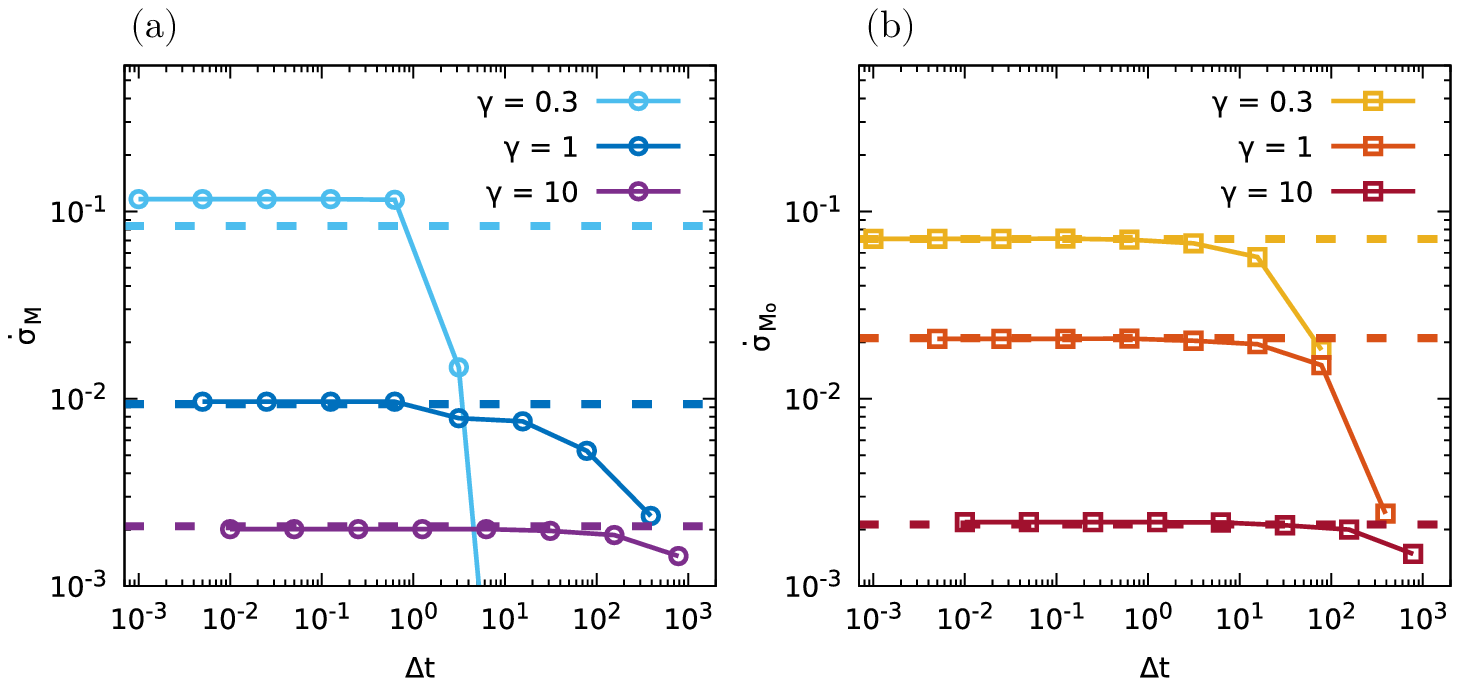}
\caption{Coarse-graining ii): continuous $\to$ discrete. 
Entropy production defined by Eq.~\eqref{eq:entropyprod_discrete} as a function of the Markov chain time-step $\Delta t$. 
Panel (a) and (b) show $\dot{\sigma}^M$ (underdamped) and $\dot{\sigma}^M_o$ (overdamped) where each curve is obtained for different values of viscosity $\gamma$.
The Markov chain (or the master equation) transition rate probabilities (to calculate $\dot{\sigma}^M$ and $\dot{\sigma}^M_o$) are obtained from the underdamped (Eq.~\eqref{eq:sigma_underdamped}) and overdamped (Eq.~\eqref{eq:sigma_overdamped}) continuous dynamics in panels~(a) and~(b), respectively.
Finally, the dashed-dotted lines mark the value of $\dot{\sigma}$ and $\dot{\sigma}_o$ in each case, maintaining the same color legend.
The first $\Delta t$ value of each panel corresponds to the master equation approximation, so that $\Delta t$ coincides with the time-step of the continous-time simulation.
The other paramters of the simulations are: $L=2\pi$, $E=1$, $A=4$, $T=1$ and $m=1$.
}
\label{fig:discrete_EP}
\end{figure*}

As a further level of coarse-graining, one can take advantage of the periodic structure of the potential to associate a suitable spatial discretization to the continuous particle position, $x(t)$.
As already shown in Fig.~\ref{fig:traj}, the particle spends much of the time in the neighborhood of the wells and, rarely, jumps forward or backward towards the neighboring minima except in the case of underdamped dynamics with small values of $\gamma$.
For this reason, a natural choice to discretize the dynamics is to map $x(t)$ onto a set of states $\{ \omega_t \} = i$ that identify
the minima of the potential where the particle is placed at time $t$.
In formulae, the particle is in the state $\omega=i$ when $x(t) \sim x_i^{min}$, being $x_i^{min}$ the position of the $i$-th minimum of the potential.
We also introduce a temporal discretization by mapping the continous time, $t$, onto the set $t_n=n\Delta t$ with $n=0, 1, 2, ..., N$, where $t_N=N \Delta t$ corresponds to the final time at which the simulation is stopped.
The time-step, $\Delta t$, cannot be arbitrarily large for consistency with the continuous dynamics.
In particular, $\Delta t$ needs to be chosen smaller than the typical forward jump times, $\tau^+$ (or $\tau^+_o$), namely $\Delta t<10^{-1} \tau^+$.
This condition guarantees that multiple jumps are rare events in agreement with the continuous time simulations.
Chosing $\Delta t \to dt$, corresponds to mapping the stocastic process, $x(t)$, onto a master equation (namely, continuous time and discrete space), while taking $\Delta t > dt$ is equivalent of mapping $x(t)$ onto a Markov chain (namely, discrete time and discrete space).
The transition probability associated to the Markov chain (or to the master equation), so far introduced, can be directly obtained from the continous simulations by calculating:
\begin{equation}
\mathcal{P}^{\pm} = \frac{j^{\pm}}{N} \,
\label{eq:transition_rate}
\end{equation}
where $j^{+}$ and $j^{-}$ are the numbers of forward and backward jumps occurring in the whole time-interval $t_N$, which are calculated by counting the number of transitions such that $\omega_{t_{n}} > \omega_{t_{n-1}}$ and $\omega_{t_{n}}<\omega_{t_{n-1}}$.
To get correct estimates, it is essential to avoid fake jumps when the particle moves around the maxima.
To do so the mapping $x(t) \to \omega_{t_n}$ is achieved using the core sets method~\cite{bowman2013introduction, sarich2010approximation, schutte2011markov} whose details are described in Appendix~\ref{app:coresets}.
Following this technique, the jump probability (which is not reported) shows that $\mathcal{P}^{\pm}/\Delta t$ are almost independent of $\Delta t$ independently of $\gamma$, as one expects.

The entropy production associated to a Markov chain with transition probability, $\mathcal{P}^{\pm}$, can be calculated as
\begin{equation}
\label{eq:entropyprod_discrete}
\dot{\sigma}^M = \frac{1}{\Delta t} \left( \mathcal{P}^{+} - \mathcal{P}^{-} \right) \log{\frac{\mathcal{P}^+}{\mathcal{P}^-}} \,.
\end{equation}
This formula neglects the probability of forward and backward multiple jumps that we expect to be exponentially rarer if $\Delta t\ll \tau^{+}<\tau^{-}$ and, thus, negligible.
At variance with the continuous estimate for $\dot{\sigma}$ or $\dot{\sigma}_o$, Eq.~\eqref{eq:entropyprod_discrete} can be managed only if the backward jumps are numerically accessible.
This technical detail restricts the applicability of this discrete coarse-grained description to the overdamped dynamics (for any $\gamma$) and to the underdamped dynamics for large or intermediate values of $\gamma$.

\section{Numerical results: entropy production after discrete coarse-graining}\label{sec:numerical_study_discrete}

In order to understand the effect of the discrete spatio-temporal coarse-graining on the entropy production, we numerically study the definition~\eqref{eq:entropyprod_discrete}. This formula can be used after applying the discrete spatio-temporal mapping discussed in Sec.~\ref{sec:discrete} on one of the two continuous dynamics discussed so far, namely Eq.~\eqref{eq:underdamped_dynamics} and Eq.~\eqref{eq:overdamped_dynamics} for the underdamped ($\dot{\sigma}^M$) and overdamped ($\dot{\sigma}^M_o$) cases, respectively.

Figure~\ref{fig:discrete_EP}~(a) and Fig.~\ref{fig:discrete_EP}~(b) show $\dot{\sigma}^M$ and $\dot{\sigma}^M_o$, respectively, as a function of the Markov chain time-step $\Delta t$: its values range from $\Delta t \to dt$ (correponding to the Master equation) to $\Delta t$ of the order of $\tau^{+}$ (average jump time). 
In both cases, this analysis is performed by evaluating three different values of $\gamma$ which belong to regime i) and ii). 
Indeed, as we have already mentioned, the discrete coarse-graining introduced so far cannot be applied to regime iii) in the underdamped case because the spatial periodicity induced by the potential is suppressed by the interplay between inertial effects and driving force.
 Each $\dot{\sigma}^M$ and $\dot{\sigma}^M_o$ are compared with the continuous expressions for the entropy production $\dot{\sigma}$ and $\dot{\sigma}^M_o$, respectively (see the dashed lines in both the panels of Fig.~\ref{fig:discrete_EP}). 
In the regime of large $\gamma$, the discrete coarse-graining leads to fairly good results both for overdamped and underdamped cases, that, in particular, are in fair quantitative agreement until to $\Delta t \sim \tau_+ /10$.
When $\gamma$ is decreased (in particular, for $\gamma=0.3$), $\dot{\sigma}^M_o$ provides still a good estimate for $\dot{\sigma}_o$, while $\dot{\sigma}^M$ overshoots $\dot{\sigma}$, because the particle starts spending much less time in the potential minima.
Finally, we observe that for $\Delta t \sim 10^{-1}\tau_+ $ the value of $\sigma^M$ (or equivalenty $\sigma^M_o$) sharply decreases with respect to the continuous time prediction.
This is because, for that value of $\Delta t$, more than one jump could occur within the same time-interval $\Delta t$ so that Eq.~\eqref{eq:entropyprod_discrete} lose some contributions (i.e. some jump events) with respect to the continuous entropy production.

\section{Conclusions}
\label{sec:summary}

In this work, we have studied how energetics properties, such as entropy production, change when the dynamics of a reference system is approximated via different levels of coarse-graining.
As a reference case, we have considered a particle in contact with a thermal bath evolving through an underdamped stochastic evolution, namely a dynamics for its position and velocity.
The particle moves in a periodic potential and is subject to a small driving force that is responsible for steady-state entropy production.
We have considered two levels of coarse-graining: i) continuous overdamped dynamics which neglects the particle velocity; ii) discrete spatio-temporal dynamics (Master equation and Markov chain) introduced by taking advantage of the periodic structure of the potential.

The coarse-graining i) reveals that inertial effects are responsible for a loss of entropy production in a regime of very small viscosity (when the inertial time is larger than the relaxation time towards to the minima of the potential), an excess in a regime of intermediate viscosity (so that the two times are comparable). 
Instead, as expected, overdamped and underdamped descriptions provide the same entropy production for large viscous coefficients.
These observations are explained by comparing entropy production and jump time stastistics: inertial forces are able to affect the average forward jump rate but lead to multiple jumps.
Instead, the second level of coarse-graining ii) is able to reproduce the value of the entropy production obtained by the corresponding continuous-time numerical simulations, provided that the time interval of the Markov chain sampling is sufficiently lower than the forward jump time and multiple jumps are negligible.


Understanding if these conclusions still hold in systems intrinsically out of equilibrium, such as those typically studied to simulate the behavior of active matter systems~\cite{bechinger2016active}, could represent an interesting challenging problem. 
In those cases, the particles usually are subject to complex mechanisms (often with biological origin) that drive the system far from equilibrium and, even in the absence of constant forces, produce entropy by their own~\cite{fodor2016far, caprini2019entropy, dabelow2019irreversibility}.
These additional degrees of freedom (with respect to passive colloids), sometimes, cannot be experimentally accessible~\cite{skinner2021improved} demanding methods  that do not require complete knowledge of the system~\cite{lestas2010fundamental, esposito2012stochastic, bisker2017hierarchical}.
In practice, this means that performing some coarse-graining could be the only reasonable choice to calculate the energetic properties of experimental active systems.
Therefore, understanding what is the effect of the coarse-graining on the estimate of the entropy production in these cases could be particularly important~\cite{busiello2020coarse}, even if, perhaps, more involved strategies could be required to perform spatial discretization.
In this context, the energetic properties of a suitable theoretical model have been already studied in Ref.~\cite{gopal2021energetics} 
where overdamped active particle moving in a sinusoidal potential because of the interplay between driving and active force have been analyzed.
Going beyond the analysis of near-equilibrium regimes~\cite{gopal2021energetics} (taking advantage of equilibrium-like approximations~\cite{wittmann2017effective, caprini2019activity}), one might wonder what are the main changes in regimes far-from-equilibrium or which includes inertial forces. Finally, clarifying whether it is safe to perform a spatio-temporal discrete coarse-graining, such as the one employed in this paper, could be an interesting issue.


\section{Acknowledgements}
This work is part of MIUR-PRIN2017 \textit{Coarse-grained description for non-equilibrium systems and transport phenomena (CO-NEST)} whose partial financial support is acknowledged. 
The authors declare no competing interests.

\appendix


\section{Numerical details: the core-sets method}\label{app:coresets}

In this Appendix, we discuss the \textit{core sets} method employed to achieve the discretization.
To do so, we need to distinguish the particle fluctuations around the minima of the potential from the jump events. 
A simple threshold between adjacent minima may not be sufficient because the thermal noise can make the particle sway on the flat profile around the maxima. 
In other words, since the region around the maximum is not stable, a crisp partition results in a large discretization error. 
The method consists in the identification of regions of space where the particle spends most of its time, called \textit{core sets} $C_i$~\cite{bowman2013introduction, sarich2010approximation, schutte2011markov}, while the remaining space is ignored.
The discrete dynamics is then entirely described by the probability of moving from a core set to the other.

In our one-dimensional periodic system, the core sets are located in the neighborhoods of the minima.
We defined the first core set as the region below a certain threshold $h$ of the complete potential, 
$$
C_0=\{x\in(M,\,M+L) \mid U(x)-E\,x<h\} \,,
$$ 
where $M=-\arcsin(E/A)$ is the position of the closest maximum to zero.
We then define all the other core sets by periodicity, $C_i=C_0+iL,\,i\in\mathbb{Z}$. 
The smaller is the threshold, the narrower and more stable are the core sets. 
In our simulations, the core set region is determined by $h\approx1+U(x_0^{min})-E\,x_0^{min}$.
The neighborhoods of the maxima $M$ are unstable by definition but, since the swaying around them is due to the thermal noise, we expect that at lower temperatures the threshold $h$ can be safely increased.

\bibliography{bibliodef}



\end{document}